\documentclass[aps,prx,showpacs,amsmath,amssymb,twocolumn]{revtex4-2}

\usepackage{graphicx}
\usepackage{amsthm}
\usepackage{bm}
\usepackage{xcolor}
\usepackage{array}
\usepackage{multirow}
\usepackage{tabularx}
\usepackage{booktabs}
\usepackage{textcomp}
\usepackage{mathtools}
\usepackage{hyperref}
\usepackage{qcircuit}
\usepackage{bbm}
\usepackage{cancel}
\usepackage{comment}
\usepackage{physics}
\usepackage{romannum}
\usepackage{algorithm}
\usepackage{algorithmic}
\usepackage{orcidlink}

\usepackage{mathtools}
\DeclarePairedDelimiter\ceil{\lceil}{\rceil}

\usepackage{listings}
\AtBeginDocument{\pagenumbering{arabic}}

\begin{document}

\title{Quantum-Enhanced Deterministic Inference of $k$-Independent Set Instances on Neutral Atom Arrays}
\author{Juyoung Park\orcidlink{0000-0002-9751-1290}}
\author{Junwoo Jung\orcidlink{0009-0000-4310-1598}}
\author{Jaewook Ahn\orcidlink{0000-0002-5837-6372}} \email{jwahn@kaist.ac.kr}
\address{Department of Physics, KAIST, Daejeon 34141, Republic of Korea}
\date{\today}

\begin{abstract}\noindent
Noisy quantum annealing experiments on Rydberg atom arrays produce measurement outcomes that deviate from ideal distributions, complicating performance evaluation. To enable a data-driven benchmarking methodology for quantum devices that accounts for both solution quality and the classical computational cost of inference from noisy measurements, we introduce deterministic error mitigation (DEM), a shot-level inference procedure informed by experimentally characterized noise. We demonstrate this approach using the decision version of the $k$-independent set problem. Within a Hamming-shell framework, the DEM candidate volume is governed by the binary entropy of the bit-flip error rate, yielding an entropy-controlled classical postprocessing cost. Using experimental measurement data, DEM reduces postprocessing overhead relative to classical inference baselines. Numerical simulations and experimental results from neutral atom devices validate the predicted scaling with system size and error rate. These scalings indicate that one hour of classical computation on an Intel i9 processor corresponds to neutral atom experiments with up to $N=250\text{--}450$ atoms at effective error rates, enabling a direct, cost-based comparison between noisy quantum experiments and classical algorithms.
\end{abstract}

\maketitle

\section{Introduction}\noindent 
A key promise of near-term quantum optimization platforms is the preparation of low-energy states of interacting many-body Hamiltonians that encode hard combinatorial problems\textcolor{blue}~\cite{Lucas2014, Albash2018RMP, Preskill2018}. In practice, however, the performance of such experiments at the level of inferred solutions is fundamentally constrained by noise: finite coherence times, control errors, and imperfect state preparation and measurement (SPAM)~\cite{Sylvain2018PRA, WJL2019, Scholl2021Nature, Henriet2020Quantum, Morgado2021AVS}. These effects obscure the connection between the implemented quantum dynamics and the intended optimization task. Consequently, even when the prepared many-body state has substantial overlap with a target configuration, noise strongly distorts the observed bitstring distribution and suppress apparent success probabilities. Extracting feasible, high-quality candidates from experimental data therefore requires classical postprocessing, whose  computational overhead must be accounted for when making fair quantum-classical comparisons.

\begin{figure*}[t]
  \centering
  \includegraphics[width=0.95\linewidth]{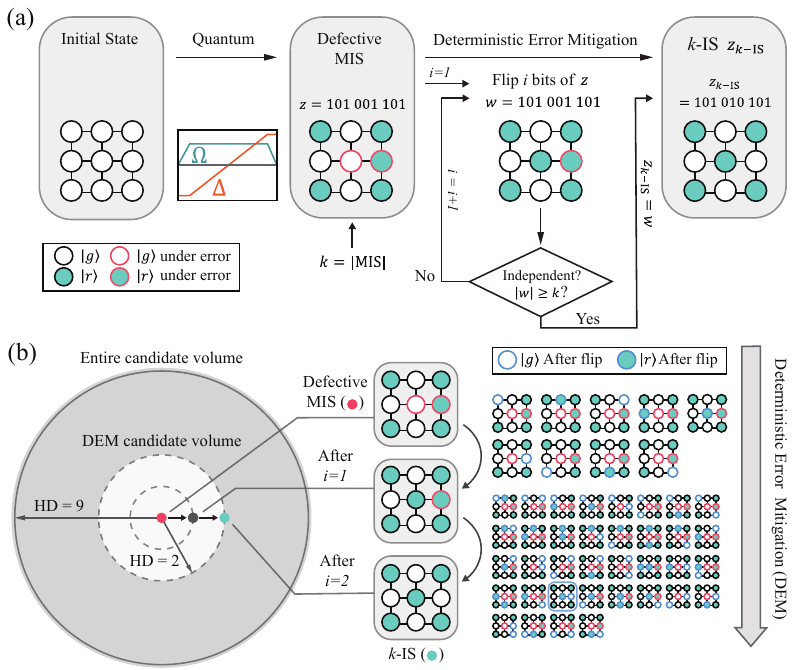}
  \caption{(a) Projective measurements of a Rydberg-atom MIS experiment produce a defective bitstring $z$ that may violate the independence constraint. Deterministic error mitigation (DEM) searches candidate bitstrings within increasing Hamming-distance shells around $z$. At distance $i$, candidate $w$ are generated by flipping $i$ bits of $z$ and checked for independence and  target size $k$. For a $3\times3$ ($N=9$ and $k=5$), the measured bitstring $z=101\,001\,101$, yields a valid solution $w = 101\,010\,101$, returning a YES instance of the $k$-independent set decision problem. (b) Geometrically, DEM expands Hamming shells around $z$, enlarging the candidate volume by $\binom{9}{1}=9$ at $i=1$ and $\binom{9}{2}=36$ at $i=2$, until a valid $k$-independent set is found.}
  \label{Fig1}
\end{figure*}

A distinctive feature of Rydberg-atom platforms~\cite{Saffman2010RMP, Pichler2018arXiv, BrowaeysLahaye2020NatPhys, Nguyen2023PRXQ, Andrist2023PRR} is that their measurement outcomes are highly structured.  Errors displace ideal configurations by several Hamming distances in a manner well approximated by a binomial distribution around an underlying solution~\cite{Byun2022PRXQ, Dalyac2023PRA, KKH2024SciData, JSH2025AQT, Dataset_JSH_2025, Tsai2025PRXQ}. Consequently, the output distribution is concentrated in a Hamming shell---the set of bitstrings at a fixed Hamming distance~\cite{CsiszarKorner2011} around the solution---with probability weight that decays rapidly as the Hamming distance increases. This structure motivates viewing error mitigation as an inference problem on a constrained and highly biased configuration manifold. From a benchmarking perspective, the Hamming shell picture also provides a natural means to quantify the associated classical postprocessing costs required to perform such inference.

A wide range of quantum error-mitigation (QEM) techniques has been developed for near-term devices. One prominent class consists of expectation-value-based statistical protocols, such as zero-noise extrapolation (ZNE)~\cite{Giurgica-Tiron2020, Mohammadipour2025, YLim2023, Raymond2025, Temme2017PRL, Endo2018PRX, Kandala2019Nature} and probabilistic error cancellation (PEC)~\cite{Nachman2020npjQI, Suzuki2022,vanDenBerg2023NatPhys,YChen2025}, which aim to reconstruct noise-free observables from noisy measurement statistics, typically at the cost of additional sampling overhead. A second class includes symmetry- and constraint-assisted techniques that suppress errors by enforcing conservation laws or projecting onto valid subspaces~\cite{Bonet-Monroig2018,Gonzalez-Garcia2025}. Both classes are naturally formulated at the level of observables or probability distributions. In contrast, quantum optimization experiments often depend on postprocessing at the level of individual measurement bitstrings, for which the associated classical computational cost is frequently implicit or omitted in performance assessments that focus primarily on solution quality.

In this context, a complementary and relevant direction is measurement-bitstring-level error mitigation, which postprocesses individual measurement outcomes to recover feasible or higher-quality configurations~\cite{Bravyi2021PRA, Funcke2022PRA, Cai2023RMP}.
Representative examples include greedy flip-based repair strategies in gate-model experiments~\cite{Dupont2023} and orijection-based repairs to feasible projections in Rydberg atom maximum-independent-set (MIS) experiments~\cite{Correc2025,Ebadi2022Science}.
These approaches can substantially improve solution quality with minimal additional hardware cost, and recent hybrid quantum-classical workflows that use measured configurations as seeds for classical simulated annealing have demonstrated notable performance gains~\cite{Wurtz2024, JSH2025AQT, Dataset_JSH_2025, Wybo2026}. Despite this practical success, the classical postprocessing cost induced by noisy samples---namely, how noise reshapes the effective candidate space and how this reshaping translates into classical inference cost---remains only partially characterized.

As device sizes increase and effective noise levels grow, lightweight local repair strategies may no longer suffice to reliably infer high-quality solutions from experimental measurements.
This motivates deterministic error mitigation (DEM), in which each measured bitstring is processed through a structured, model-informed search that systematically explores candidate configurations consistent with the device's noisy output distribution. Unlike heuristic repair schemes~\cite{Ebadi2022Science} or stochastic resampling approaches~\cite{JSH2025AQT}, DEM admits an analytic characterization: each noisy sample defines a Hamming distance shell that delineates a well-structured search space~\cite{CoverThomas1991}. This framework establishes a direct link between noise-induced Hamming structure, classical inference complexity, and the effective computational cost of postprocessing experimental outputs.

An analytic DEM model enables hardware-relevant cost accounting for quantum optimization experiments. By modeling the noisy output as a Hamming shell surrounding an underlying configuration, closed-form estimates can be derived for the dominant DEM search volume, directly linking measured error rates to the classical postprocessing cost. This perspective parallels recent approaches that benchmark quantum devices against classical simulations using empirical runtime models~\cite{Vovrosh-Dalyac2025}, while specifically targeting combinatorial optimization and highlighting the inference cost implied by experimental bitstrings. The overarching aim is to place noisy quantum dynamics and the requisite classical postprocessing on a common computational footing.
In this work, we develop an analytical and experimental characterization of the classical postprocessing cost associated with deterministic error mitigation (DEM) in independent set experiments on neutral atom quantum processors. We focus on the decision version of the $k$-independent set problem, which is NP-complete~\cite{GareyJohnson1990}, and in particular on the threshold case $k=|\mathrm{MIS}|$, corresponding to the hardest instances of the problem~\cite{Fomin2009}. Our main result is an explicit cost model controlled by binary entropy, which connects an experimentally observed Hamming-shell structure of measurement outcomes to the size of DEM candidate search space.

\section{Deterministic Error Mitigation}\label{SecII}\noindent
We develop a quantitative description of the classical computational cost of deterministic error mitigation (DEM) applied to the decision version of the $k$-independent set ($k$-IS) problem~\cite{GareyJohnson1990}. Given a graph $G$ and target size $k$, the task is to decide whether an independent set of size at least 
$k$ exists. In noisy neutral atom quantum annealing experiments, measured bitstrings may violate the independence constraint or underestimate the set size. DEM classically postprocesses individual measurement outcomes to infer whether a valid size-$k$ independent set consistent with the observed sample exists. We focus on the threshold case $k=|\mathrm{MIS}|$ and assume a noise model that captures the Hamming-shell structure of the output distribution. Within this model, we estimate the dominant DEM search volume as a function of system size 
$N$ and effective error rates, including both symmetric $P(0|1)=P(1|0)$ and asymmetric $P(0|1)\neq P(1|0)$. The resulting cost model provides a classical reference for postprocessed quantum outputs and enables comparison of experimental results.

To model noise and Hamming-distance structure, we consider an $N$-vertex graph $G(V,E)$ with an MIS bitstring $z_{\mathrm{MIS}}\in\{0,1\}^N$ and corresponding computational basis state
\begin{equation}
\ket{z_{\mathrm{MIS}}} = 
\ket{(z_{\mathrm{MIS}}^{(1)},\ldots,z_{\mathrm{MIS}}^{(N)})}, 
\end{equation}
where $\ket{0}$ and $\ket{1}$ denote the atomic ground state $\ket{g}$ and the Rydberg excited state $\ket{r}$, respectively, with $z_{\rm MIS}^{(i)}=1$ indicating that node $i$ belongs to the MIS. A noisy experimental shot yields a measured bitstring $\ket{z}=\ket{(z^{(1)},\ldots,z^{(N)})}$ via projective measurement after the quantum annealing sequence. As a minimal noise model, we assume independent binary symmetric bit-flip errors with probability $p$ on each site~\cite{CoverThomas1991}, often referred to as a SPAM error rate~\cite{Sylvain2018PRA,WJL2019}. The Hamming distance between a measured configuration and the ideal MIS configuration
\begin{equation}
HD(z,z_{\mathrm{MIS}})=\sum_{i=1}^N |z^{(i)} - z^{(i)}_{\mathrm{MIS}}|
\end{equation}
then follows a binomial distribution with mean $Np$ and variance $Np(1-p)$. As a result, experimentally sampled configurations are statistically concentrated within a Hamming shell around $z_{\mathrm{MIS}}$, rather than being uniformly distributed over the full configuration space.

To quantify the classical postprocessing cost of inferring the $k$-independent set decision problem, we first consider a brute-force deterministic error mitigation (BF-DEM) procedure. The procedure is schematically illustrated in Fig.~\ref{Fig1}(a) and described in detail in Algorithm~\ref{alg:bf_dem} (below). This BF-DEM systematically searches candidate configurations at increasing Hamming distance from a measured bitstring, serving as a minimal and transparent instantiation of the DEM framework, designed to exhaustively explore the candidate volume implied by experimentally accessible Hamming shells, rather than to optimize computational efficiency through more elaborate algorithmic structures.

\begin{algorithm}[H]
\caption{Brute-force DEM (BF-DEM)}
\label{alg:bf_dem}
\begin{algorithmic}[1]
\STATE \textbf{Input:} Graph $G(V,E)$, target solution size $k$,
\STATE \quad measured bitstring $z \in \{0,1\}^{|V|}$
\STATE \textbf{Output:} output bitstring $w$
\STATE $w \leftarrow z$
\FOR{$i \leftarrow 1$ \TO $2^{|V|}$}
  \STATE $e_i \leftarrow$ $i$-th bitstring in increasing Hamming distance order from $0^N$
  \STATE $w'\leftarrow z \oplus e_i$
  \IF{$w'$ is independent on $G$ and $|w'|>|z|$}
    \STATE $w \leftarrow w'$
  \ENDIF
  \IF{$|w| \ge k$}
    \STATE \textbf{break}
  \ENDIF
\ENDFOR
\STATE \textbf{return} $w$
\end{algorithmic}
\end{algorithm}

BF-DEM enumerates candidate configurations in order of increasing Hamming distances from the measured output $z$, as illustrated in Fig.~\ref{Fig1}(b). In the worst case, the procedure could explore nearly the entire configuration space of size $2^N$. On average, the search is restricted to a Hamming ball whose radius is set by the Hamming distance determined by the independent bit-flip error rate $p$. Since $HD(z,z_{\mathrm{MIS}})$ is typically of order $Np$, the expected number of candidates explored is well approximated by the cumulative binomial sum
\begin{align}
T(N,p)&=\sum_{k=0}^{\ceil{Np}}\binom{N}{k}=\,2^N  I_{1/2}(N-\ceil{Np},\,\ceil{Np}+1) \nonumber\\
&\approx
\frac{2^{N H_2(p)}}{\sqrt{2\pi N p(1-p)}},
\label{eqn:T(N,p)}
\end{align}
where $I_{1/2}$ denotes the regularized incomplete beta function and
\begin{equation}
H_2(p)=-p\log_2 p-(1-p)\log_2(1-p)
\end{equation}
is the binary entropy. Equation~\eqref{eqn:T(N,p)} shows that the dominant exponential scaling of the DEM search space is governed by an effective base $2^{H_2(p)}<2$ (equivalently, $T(N,p)\propto (2^{H_2(p)})^N$), corresponding to the volume of a Hamming ball in coding theory~\cite{CoverThomas1991, CsiszarKorner2011}. As $p$ increases from $0$ to $1/2$, $H_2(p)$ increases monotonically from $0$ to $1$, interpolating between a vanishing search space and full brute-force enumeration.

In realistic experiments, the readout channel is asymmetric, with distinct conditional error probabilities $p_{01}=P(1|0)$ and $p_{10}=P(0|1)$.
Let $f_0$ and $f_1$ denote the fractions of $\ket{0}$ and $\ket{1}$ sites in the ideal $k$-IS configuration. The total number of flipped bits is then the sum of two independent binomial variables, and both the mean and variance of $HD$ remain linear in $N$. To connect this asymmetric model to the symmetric expression in Eq.~\eqref{eqn:T(N,p)}, we define an effective error rate $p_{\mathrm{eff}}$ through the entropy-matching condition
\begin{equation}
H_2(p_{\mathrm{eff}})
=
f_0 H_2(p_{01}) + f_1 H_2(p_{10}).
\end{equation}
For weak asymmetry, $p_{\mathrm{eff}}$ coincides with the weighted mean error rate up to second order in $|p_{01}-p_{10}|$, so the dominant DEM scaling is governed by the same entropy-controlled form $T(N,p_{\mathrm{eff}})\propto 2^{N H_2(p_{\mathrm{eff}})}$. This relationship is confirmed in Fig.~\ref{Fig2}(c), where numerically emulated data of varying $p_{10}$ and $p_{01}$ (box charts) are compared with the theoretical prediction based on $p_{\mathrm{eff}}$, showing good agreement. Henceforth, we write $T(N,p_{01},p_{10}) := T(N,p_{\mathrm{eff}})$ in contexts where the dependence on $(p_{01},p_{10})$ is implicit through $p_{\mathrm{eff}}$.

\section{Experimental Demonstration of Deterministic Error Mitigation}\noindent
We now benchmark deterministic error mitigation (DEM) using experimental measurement bitstrings obtained from a neutral atom quantum processor, focusing on the number of classical postprocessing operations implied by the experiments. The goal is to test whether the classical costs required by DEM, as observed in experimental data, are consistent with the theoretical model established in the previous section, and to clarify the associated computational cost requirements in comparison to purely classical baselines. Accordingly, the measured classical postprocessing costs, quantified by the number of operations as a function of system size $N$, are compared to the analytical scaling relation $T(N,p_{\rm eff}) \propto 2^{N H_2(p_{\rm eff})}$.

\begin{figure*}[htbp]
\centering
\includegraphics[width=0.95\textwidth]{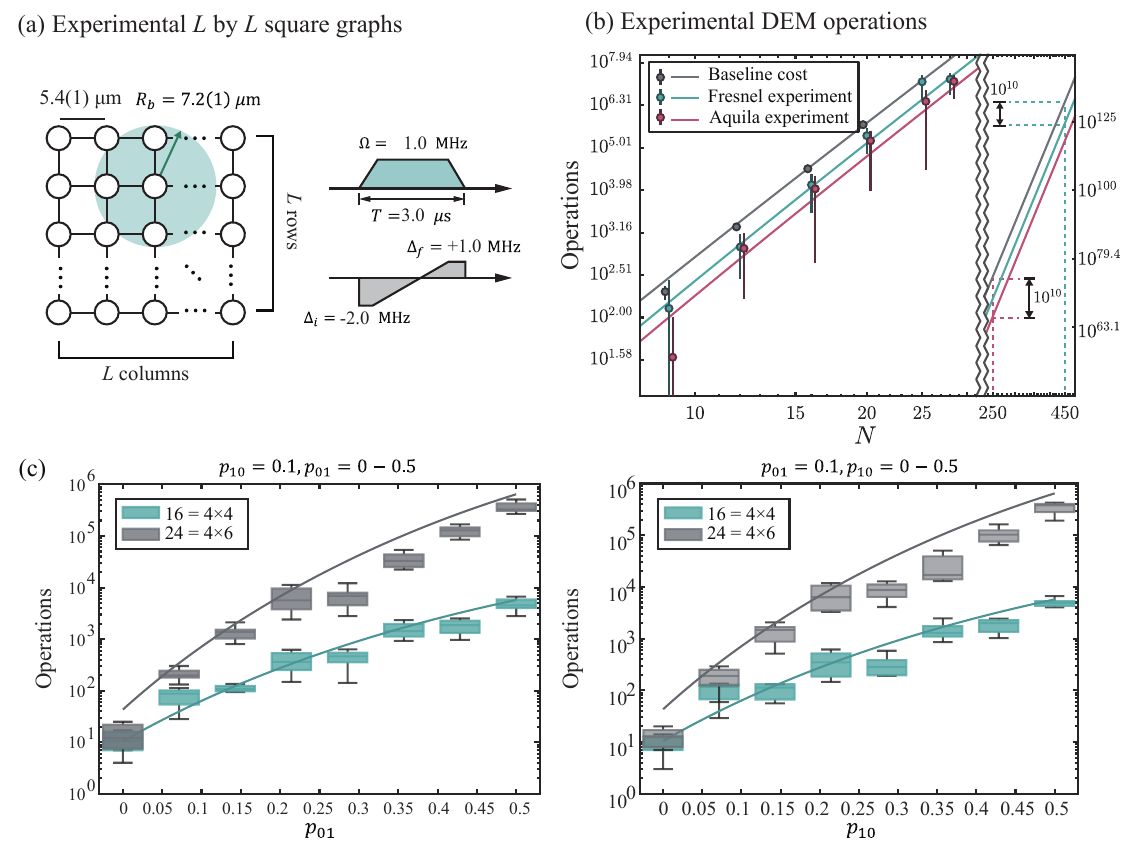}
\caption{(a) Geometry of experimentally realized $L\times L$ square Rydberg atom arrays with spacing $5.4(1)$~$\mu$m and blockade radius $R_b=7.2(1)$~$\mu$m. The MIS experiment uses a constant Rabi frequency $\Omega=1.0$~MHz and a linear detuning sweep from $\Delta_i=-2.0$~MHz to $\Delta_f=+1.0$~MHz over $T=3.0~\mu$s. (b) Measured DEM postprocessing cost versus system size $N=L^2$. Colored markers show costs from experimental data (Pasqal Fresnel, QuEra Aquila), grey markers the classical baseline, and solid lines the theoretical cost prediction from calibrated bit-flip error rates. (c) Emulated DEM postprocessing cost versus asymmetric bit-flip error probabilities $p_{01}$ and $p_{10}$ for $4\times 4$ and $4 \times 6$ lattices. Box plots show numerical emulation results, with solid lines indicating the analytic cost model $T(N,p_{01},p_{10})$. }
\label{Fig2}
\end{figure*}

The experimental setup and parameters used in this work are as follows. We used neutral atom arrays consisting of Rubidium atoms ($^{87}\mathrm{Rb}$) placed in square lattices of size $3\times 3$, $3\times 4$, $4 \times 4$, $4\times 5$, $5\times 5$, and $4\times 7$ which are illustrated in Fig.~\ref{Fig2}(a). They are realized on Pasqal Fresnel machine~\cite{Arcangelo2024PRR} and QuEra Aquila machine~\cite{Wurtz2023Arxiv}, which operate under comparable experimental settings. On the Fresnel machine, atoms are laser cooled to $20~\mu$K in a magneto-optical trap and individually trapped using optical tweezers controlled by a spatial light modulator. Each atom is optically pumped to the hyperfine ground state $\ket{0}=\ket{5S_{1/2},F=2,m_F=2}$ with a state-preparation error $\eta=0.01(1)$, and subsequently coupled to the Rydberg state $\ket{1}=\ket{60S_{1/2},m_J=1/2}$. The adiabatic pulse sequence used for MIS-type experiments consists of a constant Rabi drive $\Omega=1.0$~MHz and a linear detuning sweep from $\Delta_i=-2.0$~MHz to $\Delta_f=+1.0$~MHz over a duration of $T=3.0~\mu$s. Atoms are arranged with lattice spacing $a=5.4(1)\,\mu$m, such that nearest neighbors lie within the Rydberg blockade radius $R_b=7.2(1)\,\mu$m, while next-nearest neighbors lie outside $R_b$. This geometry enforces the Rydberg blockade constraint corresponding to MIS on the square lattice. The QuEra Aquila machine implements a similar experimental setup, differing primarily in its use of the $70S_{1/2}$ Rydberg state.

We applied the BF-DEM procedure described in Sec.~\ref{SecII} directly to experimental measurement bitstrings, using square-lattice atom arrays as illustrated in Fig.~\ref{Fig2}(a). The green (red) error-bar plots in Fig.~\ref{Fig2}(b) show the number of DEM operations extracted from Fresnel (Aquila) experimental measurements as a function of  graph sizes $N$, with gray error bars included for comparison to a purely classical brute-force solver. The $y$-axis ticks are equally spaced on a double-logarithmic scale, while the $x$-axis ticks are equally spaced on a logarithmic scale, highlighting the exponential dependence of the DEM operation count on system size $N$. The observed exponential growth with $N$ is consistent with the binary-entropy–controlled scaling predicted by the analytic model, yielding fitted effective symmetric bit-flip parameter of $p_{\rm fit} \approx 0.35$ for Fresnel and $p_{\rm fit} \approx 0.30$ for Aquila, in agreement with the independently calibrated experimental bit-flip error rates (see Appendix~\ref{app:calibration}).

Taken together, these results show that the classical cost of deterministic error mitigation on current Rydberg platforms scales in accordance with the entropy-controlled Hamming-shell model, with the leading exponential dependence on system size governed by the effective error rate through $H_2(p_{\mathrm{eff}})$. Moreover, performing DEM searches on experimentally measured bitstrings yields, on average, lower the postprocessing costs than purely classical baselines that do not leverage experimental data. These findings establish DEM as a natural reference framework for quantifying and comparing the classical postprocessing costs implied by noisy quantum optimization outputs.

So far, we have shown that the leading classical postprocessing cost is set by the volume of a Hamming ball, which scales as $2^{N H_2(p)}$ up to subexponential factors. We now place this scaling in context by comparing the postprocessing overhead implied by experimental measurements to a classical computational baseline. As a reference point, we consider a one-hour runtime on a modern desktop CPU (Intel i9), corresponding to approximately $10^{10}$ elementary operations.  Using theoretical extrapolations of the experimentally measured operation counts in Fig.~\ref{Fig2}(b), we estimate the neutral-atom system size $N$ at which the reduction in postprocessing cost relative to a classical baseline, $T(N,0.5)-T(N,p_{\rm eff})=10^{10}$, matches this computational budget. This yields characteristic system sizes of $N \approx 450(280)$ for an effective error rate $p_{\rm eff}=0.36(11)$ and $N \approx 250(150)$ for $p_{\rm eff}=0.30(13)$.  These estimates indicates that the present-day neutral atom devices can exceed the classical BF-DEM-based postprocessing capacity of contemporary desktop CPUs with only a few hundred atoms.

\section{Discussion}\noindent
We have used BF-DEM as a primary, device-calibrated reference for the DEM framework, to find that the leading asymptotic classical postprocessing cost is 
governed by the entropy-defined Hamming-ball volume implied by the measurement outcomes. A natural question is whether this  bahavior is specific to brute-force enumeration in BF-DEM. To test that this scaling persists for more efficient algorithmic backends, we consider an improved pruning-aware implementation, termed ``Tarjan-DEM'' which incorporates an exact MIS search algorithm based on standard graph-theoretic pruning~\cite{TarjanTrojanowski1977}. The Tarjan-DEM algorithm pseudocode is specified in the Algorithm~\ref{alg:tarjan_dem} (below).
\begin{algorithm}[H]
\caption{Tarjan-DEM (Entropy-informed pruning)}
\label{alg:tarjan_dem}
\begin{algorithmic}[1]
\STATE \textbf{Input:} Graph $G=(V,E)$, effective bit-flip error rate $p$
\STATE \textbf{Output:} pruned MIS size $S^\star_{\mathrm{DEM}}$
\STATE $H_2(p) \leftarrow -p \log_2 p - (1-p) \log_2(1-p)$
\STATE $K \leftarrow \lceil |V| \cdot H_2(p)\rceil$ \COMMENT{Entropy-informed pruning budget}
\STATE \textbf{Function} $\text{Solve-DEM}(G, V_{\mathrm{curr}}, \mathrm{depth})$
  \IF{$V_{\mathrm{curr}}=\emptyset$ or $\mathrm{depth} \ge K$}
    \STATE \textbf{return} $0$ \COMMENT{Subspace pruning based on entropy constraint}
  \ENDIF
  \STATE Select $v \in V_{\mathrm{curr}}$ following Tarjan's max-degree rule
  \STATE $S_{\mathrm{in}} \leftarrow 1 + \text{Solve-DEM}(G, V_{\mathrm{curr}} \setminus (\{v\} \cup N(v)), \mathrm{depth}+1)$
  \STATE $S_{\mathrm{out}} \leftarrow \text{Solve-DEM}(G, V_{\mathrm{curr}} \setminus \{v\}, \mathrm{depth})$
  \STATE \textbf{return} $\max(S_{\mathrm{in}}, S_{\mathrm{out}})$
\STATE \textbf{End Function}
\STATE \textbf{return} $\text{Solve-DEM}(G, V, 0)$
\end{algorithmic}
\end{algorithm}

We have applied this Tarjan-DEM to 2D square-lattice MIS instances of size $L\times L$ with $L=4,\dots,8$ ($N=L^2$), to generate reference MIS configurations $x^*$ and to emulate noisy measurement outcomes $z_0$ via an independent symmetric bit-flip channel with error rate $p$, i.e., $P(z_0[i]\neq x^*[i])=p$. The algorithm restricts its search to the DEM-implied neighborhood around $z_0$, namely a Hamming ball whose typical radius scales as $N_p$. This construction makes explicit the operational role of DEM: a measured bitstring defines a statistically restricted subspace of volume scaling as $2^{NH_2(p)}$ within which deterministic classical inference is performed. 

\begin{figure*}[htbp]
\centering
\includegraphics[width=0.95\textwidth]{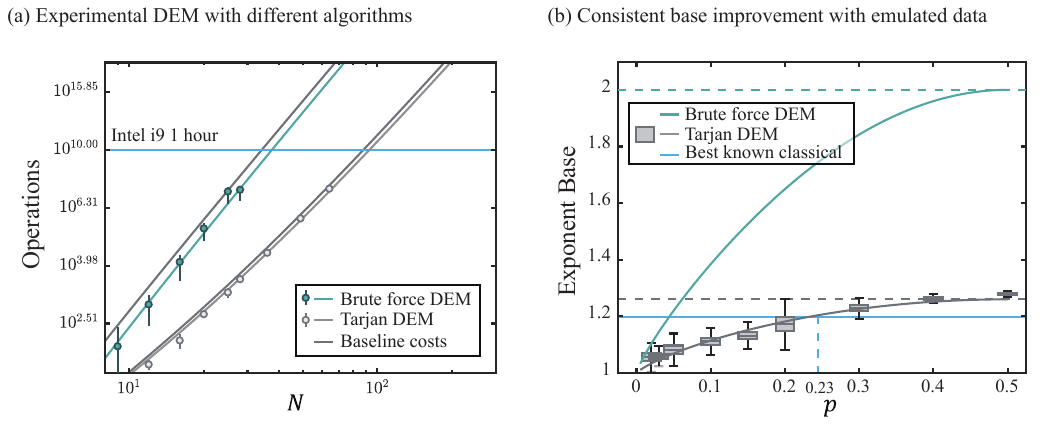}
\caption{(a) 
Operation counts for BF-DEM and Tarjan-DEM versus graph size $N$, measured on Pasqal's Fresnel device. Tarjan-DEM (light gray markers and line) shows substantially improved scaling compared to BF-DEM.
(b) 
The empirical exponent base $c$, defined by $\mathrm{Nodes} \propto c^N$, is plotted as a function of the effective error rate $p$ for 2D square lattices ($L\times L$ with $L=4,\dots,8$). The navy dashed curve indicates the BF-DEM reference, $c_{\mathrm{BF}}(p)=2^{H_2(p)}$; the black dashed curve shows the pruning-aware interpolation $c_{\mathrm{th}}(p)=c_0^{H_2(p)}$ with $c_0 \equiv 2^{1/3}$. The gray solid line is the fitted Tarjan-DEM base $c_{\mathrm{emp}}(p)$ extracted from the branch-and-reduce search tree. }
\label{Fig4}
\end{figure*}

Figure~\ref{Fig4}(a) compares the postprocessing operation counts of Tarjan-DEM and BF-DEM applied to experimental bitstrings from Pasqal Fresnel device. The postprocessing cost is quantified by the number of accounted logical branchings in the search tree of size K~\cite{FominKratsch2010,Fomin2009,TarjanTrojanowski1977}. For each $p$, we have fitted the median cost across random instances to
\begin{equation}
    \ln K(N,p) \approx N \ln c_{\text{emp}}(p) + \text{const.},
\end{equation}
so that $\ln c_{\text{emp}}(p)$ is the effective slope shown in Fig.~\ref{Fig4}(a).
This representation isolates the exponential component of the runtime and allows a direct comparison to the BF-DEM reference base $2^{H_2(p)}$. Figure~\ref{Fig4}(b) shows that $c_{\mathrm{emp}}(p)$ is captured by the interpolation
\begin{equation}
c_{\mathrm{th}}(p)=c_0^{H_2(p)}, \qquad c_0 \equiv 2^{1/3},
\end{equation}
over the full range of $p$ explored. This provides an interpretation that the DEM constraint supplies the entropy exponent $H_2(p)$, which reduces the effective search exponent from $N$ to $N H_2(p)$. The role of algorithmic pruning is that the branch-and-reduce pruning steps supply a reduced intrinsic branching base $c_0<2$ in the maximally uninformative limit $p=1/2$.
In contrast, the BF-DEM reference corresponds to the same entropy exponent but with the brute-force enumeration base $2$. Although the overall cost of Tarjan-DEM is improved, with cost exponent bases that remain below those of BF-DEM across error rates $p\in[0,0.5]$, the overall picture persists. So, the operational-cost scaling behavior dependent on the entropy-defined Hamming-ball volume is not limited to BF-DEM. So, the operational-cost scaling behavior dependent on the entropy-defined Hamming-ball volume is not specific to BF-DEM and may persist for more efficient algorithms.


\begin{figure*}[htbp]
\centering
\includegraphics[width=0.95\textwidth]{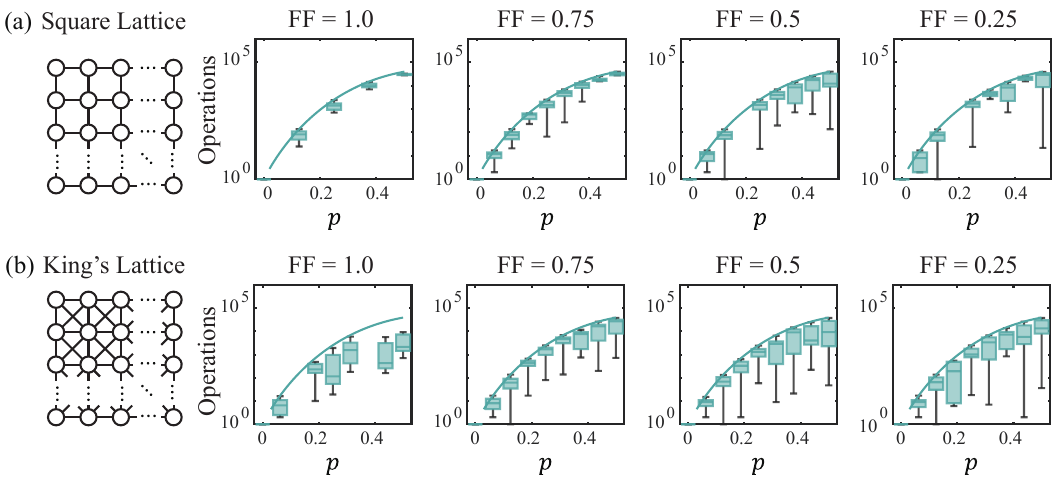}
\caption{
(a) DEM operation cost versus symmetric bit-flip error rate $p$ for MIS instances on square lattices at different filling fractions (FF), realized by embedding a fixed $16$-vertex subgraph into larger lattices. Box charts show emulated operation costs, and solid lines show theoretical predictions. (b) DEM cost for King's-lattice graphs at different filling fractions. The same dependence on $p$ persists, indicating that DEM cost is governed primarily by the effective error rate rather than detailed graph geometry.}
\label{Fig3}
\end{figure*}

Another natural question is whether the scaling behavior of DEM depends on the underlying graph geometry, or whether it persists regardless of the connectivity and filling fraction of the graph. Figures~\ref{Fig3}(a,b) address this question by examining the DEM cost as a function of the
bit-flip error rate for different lattice geometries—square lattices in (a) and King's lattices in (b)—and for different filling fractions (FF). Different filling fractions are generated by embedding the same $16$-vertex subgraph into lattices of increasing size, which changes the local connectivity and vertex density while preserving the system size. Across both square and King's lattice geometries, the DEM cost exhibits the same qualitative dependence on the error rate. For a fixed bit-flip error rate $p$, the cost grows along the theoretical prediction across all filling fractions, despite differences in graph structure. This indicates that variations in lattice geometry have only a minimal effect on the leading scaling of the DEM cost. Therefore, the growth of the DEM cost with increasing error rate is controlled by the size of the Hamming shell defined by the noise level, rather than by microscopic details of the graph.

Lastly, our positioning of deterministic error mitigation (DEM) has thus far relied on comparisons with restricted classical baselines. A natural counterargument is that the appropriate reference should instead be the best polynomial-time classical algorithm~\cite{Ronnow2014Science}, for example by generating seed candidate bitstrings classically and then applying DEM as a postprocessing step. However, there is no established one-to-one mapping between quantum experimental operations and classical computational steps, nor a definitive notion of a “best” classical algorithm for NP-hard optimization problems~\cite{Wolpert1997IEEE}. Nevertheless, meaningful comparisons can be made using standard complexity-theoretic bounds. Approximating the maximum independent set of an $N$-vertex graph beyond an approximation ratio $\alpha \gtrsim 0.73$ is NP-hard~\cite{Hastad2001,DinurSafra2005}, which for square lattice corresponds to an effective bit-flip error rate of $p \simeq 0.27$. This sets an upper bound on the quality of seed configurations achievable by polynomial-time classical algorithms. Absent detailed geometric structure, the effective error rate of  classically generated seeds is expected to approach the uninformative limit  $p=0.5$ as $N$ grows. In our entropy-controlled model, this implies a DEM search volume approaching the full configuration space, with postprocessing cost scaling as the classical exponential baseline, $T(N,p=0.5)=O(2^N)$ for BF-DEM and $O(2^{N/3})$ for Tarjan-DEM. Thus, while small systems may admit stronger classical baselines, $p=0.5$ dominates asymptotically, and experimentally produced bitstrings with $p_{\mathrm{eff}}<0.27$ represent a genuine resource for reducing inference cost.

\section{Conclusion}\noindent
We have developed a deterministic error mitigation (DEM) framework for the decision version of the $k$-independent set problem in MIS-type neutral atom quantum annealing experiments, with an explicit focus on quantifying the classical postprocessing cost implied by noisy experimental outputs. By formulating bitstring-level error mitigation as a deterministic inference task constrained by an experimentally motivated Hamming structure, we derived an entropy-controlled model for the dominant DEM candidate volume and its associated classical postprocessing cost. The resulting scaling is validated by numerical emulations and by experimental cost trends observed on contemporary Rydberg-atom quantum processors. Beyond quantitative estimates, the framework isolates the leading scaling of postprocessing cost from solver- and implementation-specific details. This enables hardware-relevant comparisons between postprocessed quantum optimization outputs and classical baselines, placing solution quality and classical inference cost on a common computational footing.



\section*{Acknowledgements}\noindent
This research was supported by National Research Foundation of Korea (NRF) under Grant No. RS-2024-00340652 and RS-2025-25464441, and by the Institute of Information \& Communications Technology Planning \& Evaluation (IITP) under
Grant No. IITP-2025-RS-2024-00437191, funded by the Korea government (MSIT). The authors acknowledge the use of the Pasqal Fresnel hardware via the Pasqal Cloud service and the QuEra Aquila hardware via the Amazon Braket service in obtaining the experimental results reported here. We also acknowledge Changhun Oh for fruitful discussions. 



\appendix
\section{Calibration of experimental bit-flip error rates}\label{app:calibration}\noindent
We compare independently calibrated bit-flip error rates in neutral atom experiments with the effective error rates inferred from DEM scaling to validate the fitted parameters. For Pasqal Fresnel, comparing ideal MIS solutions with measured bitstrings for graphs of size $N=16-100$ yields asymmetric rates $p_{10}=0.40(7)$ and $p_{01}=0.33(9)$, corresponding to $p_{\rm eff}=0.36(11)$. In contrast, single-atom SPAM calibrations give smaller rates $p_{10} = 0.20(11)$ and $p_{01} = 0.19(9)$, or $p_{\rm eff} = 0.20(14)$. The agreement between the many-body $p_{\rm eff}=0.36(11)$ and DEM operation-count scaling indicates that DEM cost is governed by many-body effective noise, consistent with accumulated control errors $p_{\rm control}=0.11(6)$~ \cite{KKH2024SciData}, rather than single-qubit errors alone. For QuEra Aquila, we extract $p_{10}=0.46(18)$, $p_{01}=0.13(12)$, and $p_{\rm eff}=0.30(13)$.

\section{Effective error rate for asymmetric SPAM noise}\noindent
In the main text, SPAM noise is captured by a single effective bit-flip probability controlling the entropy-based DEM scaling. In practice, readout errors are asymmetric, with $p_{01}\neq p_{10}$. We map this asymmetry to an effective symmetric error rate $p_{\rm eff}$. Let  $p_{01} = P(1|0)$, $p_{10} = P(0|1)$, and  $f_0$ ($f_1$) be the fraction of $\ket{0}$ ($\ket{1}$) sites in the ideal MIS,  with $f_0+f_1=1$. Using the binary entropy $H_2(p)$,  we obtain $p_{\rm eff}$  by equating entropy costs:
\begin{equation}
  H_2(p_{\rm eff})  =   f_0 H_2(p_{01}) + f_1 H_2(p_{10}).
  \label{eq:peff_def_appendix}
\end{equation}
Writing  $\bar p = f_0 p_{01} + f_1 p_{10}$ and $\Delta p = p_{01} - p_{10}$, a second-order Taylor expansion yields
\begin{equation}
  p_{\rm eff}
  \simeq
  \bar p
  +
  \frac{1}{2}
  \frac{H_2''(\bar p)}{H_2'(\bar p)}
  f_0 f_1 (\Delta p)^2
  +
  \mathcal{O}((\Delta p)^4),
  \label{eq:peff_second_order}
\end{equation}
with $H_2'(p) = \log_2\!\left( \frac{1-p}{p} \right)$ and $H_2''(p)= -\frac{1}{\ln 2} \frac{1}{p(1-p)}$.
Since $H_2$ is concave, the correction is negative: $p_{\rm eff} < \bar p$, with the deviation scaling as $f_0 f_1 (\Delta p)^2$ and maximized for balanced MIS configuration and large SPAM asymmetry.

\section{Tarjan-DEM time complexity under an entropy constraint.} \noindent
DEM restricts candidates to a Hamming ball $\mathcal B(z_0,r)$ around a measured bitstring $z_0$. For i.i.d. Bernoulli bit-flips (binary symmetric channel) with rate $p$~\cite{CoverThomas1991}, the typical radius is $r\simeq Np$, giving an effective search space
\begin{equation}
\begin{split}
M_{\mathrm{eff}}
:= |\mathcal B(z_0,Np)|
= \sum_{k=0}^{Np}\binom{N}{k}
\approx
\frac{2^{N H_2(p)}}{\sqrt{2\pi Np(1-p)}}.
\end{split}
\label{eq:Meff}
\end{equation}
Thus, $M_{\mathrm{eff}} \sim 2^{N H_2(p)}$ up to subexponential factors~\cite{CoverThomas1991,CsiszarKorner2011}. The Tarjan-Trojanowski MIS solver has runtime $T_{\mathrm{TT}}(N)=\tilde{\mathcal O}(c_0^N)$ with $c_0<2$~\cite{TarjanTrojanowski1977,FominKratsch2010}, or $T_{\mathrm{TT}}=\tilde{\mathcal O}(M^\gamma)$ for $M=2^N$ and $\gamma=\log_2 c_0\in(0,1)$. Restricting the search space to $M_{\rm eff}$ yields
\begin{equation}
T_{\mathrm{TT}}(N)=\tilde{\mathcal O}\!\bigl(M^{\gamma}\bigr)=\tilde{\mathcal O}\!\bigl(c_0^{N H_2(p)}\bigr).
\end{equation}
For asymmetric readout noise, $p$ is replaced by entropy-matched effective rate $p_{\mathrm{eff}}$.

\end{document}